# Bragg Coherent Modulation Imaging for Highly Strained Nanocrystals—A Numerical Study

Jiangtao Zhao[1,2], Ivan A. Vartanyants[2], and Fucai Zhang[1]*

[1]*Department of Electrical and Electronic Engineering, Southern University of Science and Technology, 1088 Xueyuan Avenue, Shenzhen 518055, China*

[2]*Deutsches Elektronen-Synchrotron DESY, Notkestraße 85, 22607 Hamburg, Germany.*

*zhangfc@sustech.edu.cn

**Abstract:**

Bragg coherent diffraction imaging (BCDI) is a unique and powerful method for tracking three-dimensional strain fields non-destructively. While BCDI has been successfully applied to many scientific research fields and receives high demands, the reconstructed results for highly strained crystals are still subject to big uncertainties. Here, the progress in improving the suitability of BCDI for general samples by exploiting wavefront modulation is reported. Extensive numerical simulations demonstrate that significant improvements over the current method for reconstructing highly strained model nanocrystals can be achieved. The proposed method highly suppresses the appearance of ambiguous solutions and exhibits fast convergence and high robustness in phase retrieval. Possible experimental difficulties in implementing this method are discussed in detail.

## 1 Introduction:

X-ray Bragg coherent diffraction imaging (BCDI) is a non-destructive method for measuring the three-dimensional (3D) lattice strain field of nanocrystals by phasing sufficiently sampled Bragg diffraction patterns (Robinson *et al.*, 2001, Robinson & Harder, 2009, Pfeifer *et al.*, 2006). The amplitude of the reconstructed 3D object function represents the crystal shape function while the phase accurately represents the projection of the lattice displacement along a particular scattering vector (Vartanyants

& Robinson, 2001, Favre-Nicolin *et al.*, 2010). The BCDI phase retrieval algorithm has been successfully applied to some mildly strained crystals (Yang *et al.*, 2013, Clark *et al.*, 2015, Karpov *et al.*, 2017, Singer *et al.*, 2018, Kawaguchi *et al.*, 2019, Carnis *et al.*, 2021, Dupraz *et al.*, 2022). However, phasing the diffraction intensity of highly strained nanocrystals encounters considerable difficulties as the resulting reconstruction could show nonphysical crystallographic electron density 'gaps' in regions where the crystal phase exceeds $\pm\pi/2$ and varies rapidly (Newton *et al.*, 2010, Huang *et al.*, 2011, Newton, 2012).

To overcome the limitations on imaging highly strained nanocrystals, several extensions to BCDI have been suggested (Minkevich *et al.*, 2008, Minkevich *et al.*, 2011, Newton *et al.*, 2010, Newton, 2012, Huang *et al.*, 2011, Wang, Gorobtsov, *et al.*, 2020, Gao *et al.*, 2021). However, these extensions may not always fit the conditions in real situations. For instance, the method provided by Minkevich *et al.* (Minkevich *et al.*, 2008, Minkevich *et al.*, 2011) introduced an extra density uniform constraint and a confined phase difference between neighboring points in the algorithm, and to reach a correct solution, *a priori* knowledge such as an accurate support size and the range of object phase should be known. The following compressive-sensing-based density modification (Newton, 2012) and two-step phase-constrained hybrid input-output (HIO) phase retrieval algorithms (Huang *et al.*, 2011) only achieved limited success in phasing highly strained crystals under the assumption of a continuous electron density distribution. Another line of thought is to increase the data redundancy for better reconstruction accuracy if some specific requirements can be fulfilled; for example, the approach proposed by Wang *et al.* (Wang, Gorobtsov, *et al.*, 2020) was under the constant electron density assumption when the studied crystals undergo phase transformation. While in the work of Gao *et al.* (Gao et al., 2021), the simultaneous inversion of multiple Bragg peaks would require careful selection of the Bragg peak pairs and thus hindered by the increased experimental difficulties (Lauraux *et al.*, 2021). Bragg ptychography (Hruszkewycz *et al.*, 2017, Li *et al.*, 2021) is a viable way of measuring highly strained samples, but, it is more suitable for single crystal thin films. At present, Bragg ptychography experiments are still facing great challenges due to their high requirements for equipment stability.

Introducing the idea of wavefront modulation into BCDI may be a great alternative approach to address these difficulties in imaging highly strained nanocrystals. The advantages of wavefront modulation have already been demonstrated in the two-dimensional (2D) transmission coherent modulation imaging (CMI) (Zhang & Rodenburg, 2010, Zhang *et al.*, 2016). The insertion of the wavefront modulator in the path between the object and detector not only strengthens phase encoding of the sample exit wave into the recorded diffraction data, but also breaks the conditions of the occurrence of ambiguous solutions. As a result, the phase retrieval exhibits a fast convergence rate for general samples. A previous report (Ulvestad *et al.*, 2018) that attempted to introduce wavefront modulation into BCDI has shown some encouraging results. However, their simulation demonstration is limited to the case of a 2D projection of crystal lattice strain. Expanding the wavefront modulation idea into 3D and a detailed investigation of its performance is thus well worth exploring.

In this work, we have implemented the wavefront modulation onto the 3D Bragg diffraction. The principles of the proposed Bragg coherent modulation imaging (BCMI) approach and its comparison with traditional BCDI are given in section II. The numerical simulations of two special models of highly strained nanocrystal, which were used to demonstrate the robustness and stability of the BCMI phase retrieval, are presented in section III. In section IV, we discuss the experimental implementation of BCMI and the associated difficulties in detail.

## 2. BCMI Principles

Before describing the BCMI approach, we first briefly review the formulation of Bragg diffraction in BCDI. The schematics of BCDI are shown in Fig. 1(a), where a monochromatic X-ray plane wave illuminates a finite nanocrystal with a periodic electron density $\rho$ and exits under the Bragg condition. The reciprocal space volume in the vicinity of the measured Bragg peak is mapped through rocking curve measurements by changing the angle $\theta$. Under the kinematic approximation, the 3D scattering amplitude $E(Q)$ of the finite nanocrystal at the vector $\boldsymbol{h} = 2\pi\boldsymbol{H}$, where $\boldsymbol{H}$ is the reciprocal lattice vector of the sample, can be expressed as (Vartanyants & Robinson, 2001, Favre-Nicolin *et al.*, 2010)

$$E(\mathbf{Q}) = \int S(\mathbf{r})\,e^{-i\mathbf{Q}\cdot\mathbf{r}}d\mathbf{r} = \int s(\mathbf{r})e^{-i\mathbf{h}\cdot\mathbf{u}(\mathbf{r})}e^{-i\mathbf{Q}\cdot\mathbf{r}}d\mathbf{r}. \tag{1}$$

Here, $\mathbf{r}$ and $\mathbf{Q}$ are real and reciprocal space coordinate vectors. $\mathbf{Q} = \mathbf{q} - \mathbf{h}$, $\mathbf{q} = \mathbf{k}_f - \mathbf{k}_i$ is the momentum transfer vector defined by the incident and exit wave vectors, $|\mathbf{k}_i| = |\mathbf{k}_f| = k = 2\pi/\lambda$, where $\lambda$ is the wavelength. The complex function $S(\mathbf{r})$ consists of the crystal shape function $s(\mathbf{r})$ and the phase $\phi(\mathbf{r}) = \mathbf{h}\cdot\mathbf{u}(\mathbf{r})$, where $\mathbf{u}(\mathbf{r})$ is the displacement field of the atomic planes perpendicular to $\mathbf{h}$. The measured 3D diffraction intensity is then expressed as $I(\mathbf{Q}) = |E(\mathbf{Q})|^2$. Equation (1) consists of the 3D Fourier transform and forms the basics of BCDI to recover the lost phase of $E(\mathbf{Q})$ and then to determine $S(\mathbf{r})$ through the inverse Fourier transform.

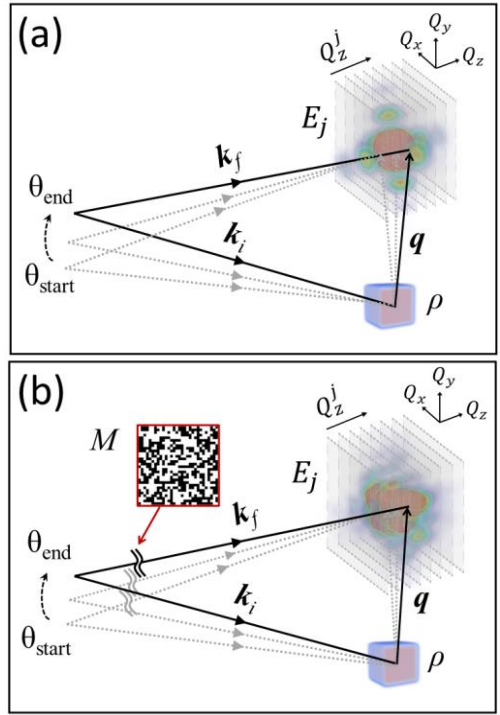

**Figure 1.** Schematic diagrams of (a) the BCDI and (b) BCMI approaches. $\mathbf{k}_i$ and $\mathbf{k}_f$ are the incident and exit wave vectors, and $\mathbf{q} = \mathbf{k}_f - \mathbf{k}_i$ is the momentum transfer vector. The orientation of $\mathbf{k}_i$, $\mathbf{k}_f$, and $\mathbf{q}$ satisfies the Bragg condition, $\rho$ is the electron density of the crystal, $\theta$ is the rocking angle and $E_j$ is the 2D measured diffraction intensity slice. $M$ is a 2D wavefront modulator.

In fact, in the case of the BCDI, the scattered amplitude $E(\mathbf{Q})$ can also be expressed in a slice-by-slice manner. The 2D cut in reciprocal space for a particular reciprocal vector $Q_z^j$ can be formulated by rearranging equation (1) as (Vartanyants & Robinson, 2001)

$$E_j(Q_x, Q_y) = \iint S_j(x, y) \, e^{-i(Q_x \cdot x + Q_y \cdot y)} dx dy, \tag{2}$$

where the subscript $j$ is the index of the reciprocal cut along the $Q_z$-direction and $S_j(x, y)$ is defined as

$$S_j(x, y) = \int S(\mathbf{r}) \, e^{-iQ_z^j \cdot z} dz. \tag{3}$$

Thus, equation (2) provides an alternative perspective to equation (1), implying that $E(\mathbf{Q})$ can be built up through a series of 2D Fourier spectrums of the phase-modulated object projections (Cha *et al.*, 2016, Li *et al.*, 2020). Similarly, the inversion of equation (1) to real space can be formulated using $E_j(Q_x, Q_y)$. The 3D real-space component $S_j(\mathbf{r})$ is related with $E_j(Q_x, Q_y)$ as

$$\begin{aligned} S_j(\mathbf{r}) &= \int E(\mathbf{Q}) \delta(Q_z - Q_z^j) e^{i\mathbf{r} \cdot \mathbf{Q}} d\mathbf{Q} \\ &= e^{iz \cdot Q_z^j} \cdot \iint E_j(Q_x, Q_y) e^{i(x \cdot Q_x + y \cdot Q_y)} dQ_x dQ_y \end{aligned}, \tag{4}$$

where $\delta(Q_z - Q_z^j)$ is a Dirac function, and the back-projection operation is implicitly performed by the sampling property of the Dirac function (Cha *et al.*, 2016, Li *et al.*, 2020). As a result, the complete real-space representation $S(\mathbf{r})$ will then be calculated by summing up all $S_j(\mathbf{r})$ components,

$$S(\mathbf{r}) = \sum_{j=1}^{J} S_j(\mathbf{r}), \tag{5}$$

where $J$ is the total number of reciprocal cuts.

The wave propagation model in the BCMI will exactly follow the slice-by-slice expression of the BCDI in equations (2)-(5) but with some modification. Fig. 1(b) shows the schematic diagram of BCMI. Compared with the conventional BCDI configuration shown in Fig. 1(a), a 2D wavefront modulator with the transmission function of $M(x, y)$ is inserted between the object and detector. This modulator is usually located a few millimeters downstream of the object and orientated parallel to the detector. It can have a random transmission or phase distribution and needs to induce significant modulation on the object exit wave. For each rocking angle, the diffracted object exit wave will travel through the modulator before reaching the detector. Thus, the BCMI propagation model involves a two-step wave propagation and a specific modulation operation in between.

In the BCMI forward propagation, the object exit wave should first be propagated to the modulator, then the modulation is applied, and finally, propagated to the detector. The wavefront propagation from the object to the modulator can be expressed as

$$E_j(x_m, y_m) = \iint P(x_m - x_o, y_m - y_o) S_j(x_o, y_o) dx_o dy_o , \qquad (6)$$

where the subscripts $m$ and $o$ specify the coordinates at the modulator and object plane, respectively. The wave propagator $P(x_m - x_o, y_m - y_o)$ has the form

$$P(x_m - x_o, y_m - y_o) = \frac{k}{i2\pi L} \exp\left[ik \frac{(x_m - x_o)^2 + (y_m - y_o)^2}{2L}\right] , \qquad (7)$$

where $L$ is the propagation distance. The modulation of $E_j(x_m, y_m)$ is expressed as

$$E_j^{out}(x_m, y_m) = M(x_m, y_m) \cdot E_j(x_m, y_m) , \qquad (8)$$

in which the modulator function is defined by

$$M(x_m, y_m) = T(x_m, y_m) e^{i\phi(x_m, y_m)} , \qquad (9)$$

where $T(x_m, y_m)$ reflects the transmission and $\phi(x_m, y_m)$ represents the phase shift introduced on $E_j(x_m, y_m)$. The $E_j(Q_x, Q_y)$ at the detector plane is finally calculated as

$$E_j(Q_x, Q_y) = \iint E_j^{out}(x_m, y_m) e^{-i(Q_x \cdot x_m + Q_y \cdot y_m)} dx_m dy_m . \qquad (10)$$

Accordingly, for the BCMI backward propagation, each $E_j(Q_x, Q_y)$ will be propagated back to the modulator, demodulated, and then propagated back to the object plane to generate the corresponding 3D real-space component $S_j(r)$. These processes are formulated as

$$E_j^{out}(x_m, y_m) = \iint E_j(Q_x, Q_y) e^{i(x_m \cdot Q_x + y_m \cdot Q_y)} dQ_x dQ_y , \qquad (11)$$

$$E_j(x_m, y_m) = M^{-1}(x_m, y_m) \cdot E_j^{out}(x_m, y_m) , \qquad (12)$$

where $M^{-1}$ is the inverse of $M$, which is

$$M^{-1}(x_m, y_m) = \frac{1}{T(x_m, y_m)} e^{-i\phi(x_m, y_m)} , \qquad (13)$$

and

$$S_j(\mathbf{r}) = e^{iz \cdot Q_z^j} \cdot \iint P^*(x_o - x_m, y_o - y_m) E_j(x_m, y_m) dx_m dy_m , \qquad (14)$$

where $P^*$ is the conjugate of $P$. The complete real-space representation $S(\mathbf{r})$ is finally calculated as formulated in equation (5). We remind that the demodulation of equation (12) suits well for a phase modulator, however, for a complex-valued modulator the expression should be the same as the decoupling of a probe function from the object exit wave in the ptychography iteration engine to avoid the zero division problem (Rodenburg & Faulkner, 2004).

With the above slice-by-slice modelling of wave propagation in BCMI, a dedicated phase retrieval algorithm was constructed (see the Appendix A for details). We next demonstrate the validity of the BCMI phase retrieval on highly strained nanocrystals through numerical simulations.

## 3. BCMI Numerical Results

Two highly strained gold (Au) nanocrystal models with the same cubic shape function but different phase distributions were used in the simulations. Figs. 2(a) and 2(b) illustrate the 3D amplitude profile and phase cross-sections in the center of the crystal. The size of the crystals was set as $510 \times 510 \times 510$ nm³ which occupied $42 \times 42 \times 42$ voxels out of a $256 \times 256 \times 256$ volume. The phase of crystal *model-1* is an isotropic type with a functional form (Newton *et al.*, 2010)

$$\phi(\mathbf{r}) = \beta \cos(0.7\pi x/\alpha_x) \cos(0.7\pi y/\alpha_y) \cos(0.7\pi z/\alpha_z) , \qquad (15)$$

and crystal *model-2* has a non-isotropic phase (Li *et al.*, 2021) defined as

$$\phi(\mathbf{r}) = \beta(x/\alpha_x) \cdot (0.8z/\alpha_z) . \qquad (16)$$

In these expressions, $\alpha$ is the object length along each orthogonal axis and $\beta$ is a factor to adjust the phase range. We used $\beta = 6$ in *model-1* and $\beta = 20$ in *model-2* where the absolute phase range is $[0, 6]$ radians and $[0, 20]$ radians, respectively.

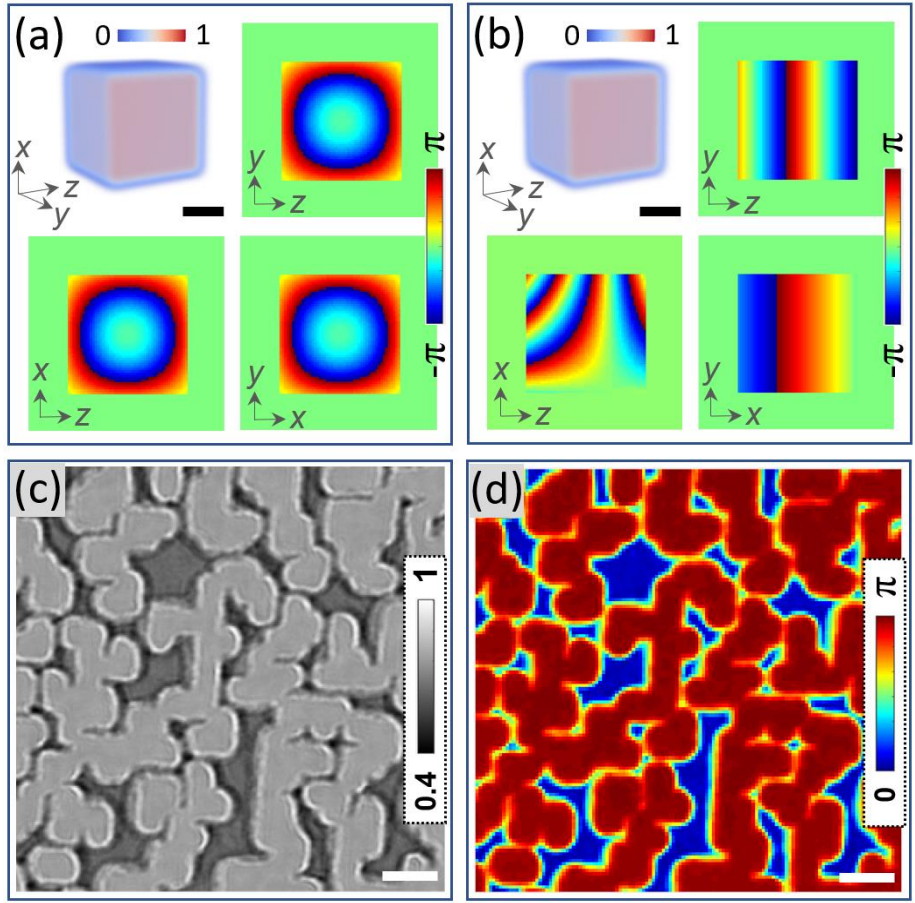

**Figure 2.** Simulated highly strained nanocrystal models and the modulator. (a), (b) The ground truth of crystal *model-1* and *model-2* presented in 3D amplitude and 2D central phase cross-sections. (c), (d) The amplitude and phase of the used modulator transmission function. The scale bar is 200 nm.

The wavefront modulator used here was a binary type, adopted from our previous study (Zhang *et al.*, 2016). This complex-valued modulator allows the scattering wave to be transmitted between $[0.4, 1]$ and holds the capability to alter the phase between $[0, \pi]$ radians for 8 keV radiation [see Figs. 2(c) and 2(d)]. The modulation units are randomly distributed and their minimum width is about 120 nm. Such a structure can guarantee an efficient modulation of the incoming diffracted wave.

The Au (111) Bragg condition ($\theta_B = 19.25°$) was considered to generate the diffraction pattern under the illumination of an 8 keV X-ray. Simulated rocking curve measurements were performed at the in-plane $\theta - 2\theta$ Bragg geometry where the separation between the object, modulator, and detector are $L_1 = 2$ mm and $L_2 = 1.5$ m [see Fig. 3(a)]. The total angular scan range was set from 18.75° to 19.75° and the scan step was $\Delta\theta = 0.01°$. We set the data acquisition window to be $256 \times 256$ pixels$^2$ and the simulated 2D detector has a pixel size ($p = p_x = p_y$) of $55 \times 55$ µm$^2$. We used a Cartesian orthogonal frame for the 3D diffraction intensity that was sampled in the $Q_x$, $Q_y$ frame with a sampling rate

$$\Delta Q_x = \Delta Q_y = \frac{2\pi}{\lambda}\frac{p}{(L_1+L_2)} = 2.02 \times 10^{-3}\ nm^{-1}, \quad (17)$$

and the sampling rate in the $Q_z$ axis is

$$\Delta Q_z = |q|\Delta\theta \cdot \cos\theta_B = \frac{4\pi}{\lambda}\sin\theta_B \cdot \Delta\theta \cdot \cos\theta_B = 4.4 \times 10^{-3}\ nm^{-1}. \quad (18)$$

Thus, the total detecting range in $Q_x$ and $Q_y$ was $[-0.258, 0.256]\ nm^{-1}$, and in $Q_z$ was $[-0.220, 0.220]\ nm^{-1}$.

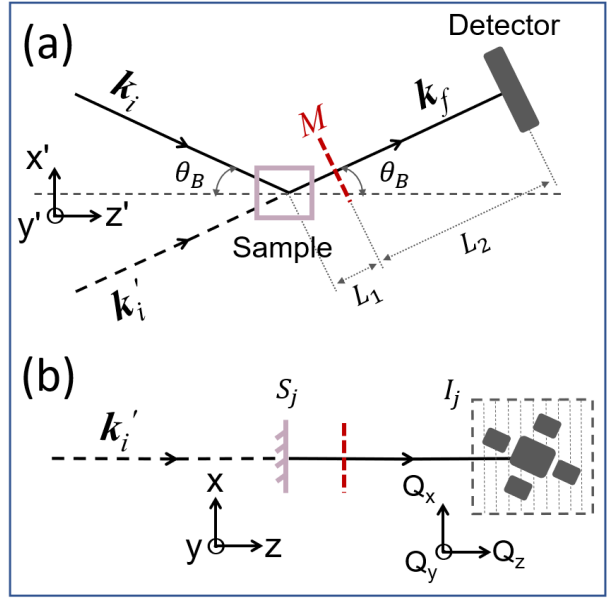

**Figure 3.** (a) The in-plane $\theta - 2\theta$ Bragg geometry used for BCMI and BCDI diffraction pattern generation (top-view from sample space). $\theta_B$ is the Bragg angle and $M$ is the modulator. The dashed line of $k_i'$ is an imaginary source of $k_i$ illumination and the $k_f$ is the exit wave. $L_1$ and $L_2$ are the separations between the object, modulator, and detector. (b) The equivalent geometry used to calculate the diffraction pattern from laboratory space to detector space. $S_j$ is the phase-modulated object projection and $I_j$ is the calculated diffraction intensity slice.

As shown in Fig. 3(b), wave propagation from laboratory space to detector space follows three steps for each rocking curve angle in generating BCMI diffraction patterns. The first step is to calculate the phase-modulated object projection $S_j$ based on equation (3), where the phase term is calculated with the above $Q_z$ value and the projecting direction is along the $z$-axis. The second step is using equation (6) through the angular spectrum method to propagate the wavefield to the modulator plane. The wavefield is then modulated following equation (8), before finally being propagated to the detector plane through a fast Fourier transform. The diffraction intensity slice $I_j$ was calculated by squaring the diffracted wave. We added Poisson noise to each of $I_j$ and stacked all the $I_j$ to get the 3D BCMI diffraction patterns. The total number of photons in 3D reciprocal space was adjusted to be about $5 \times 10^6$ which is a typical value obtainable at third-generation synchrotron sources. The 3D BCDI diffraction patterns were also generated using the same procedures above but without the second step.

Fig. 4 illustrates the details of the generated BCDI and BCMI diffraction patterns for our highly strained crystal models. Here, typical broadened BCDI diffraction patterns were obtained due to a strong phase gradient in the crystals [see Figs. 4(a) and 4(b)]. The BCMI diffraction patterns, on the other hand, have a further divergent intensity in $Q_x Q_y$ plane [see Figs. 4(c) and 4(d)], which is caused by the modulation of the wavefront modulator. This divergent diffraction intensity contains more information and can facilitate phase retrieval.

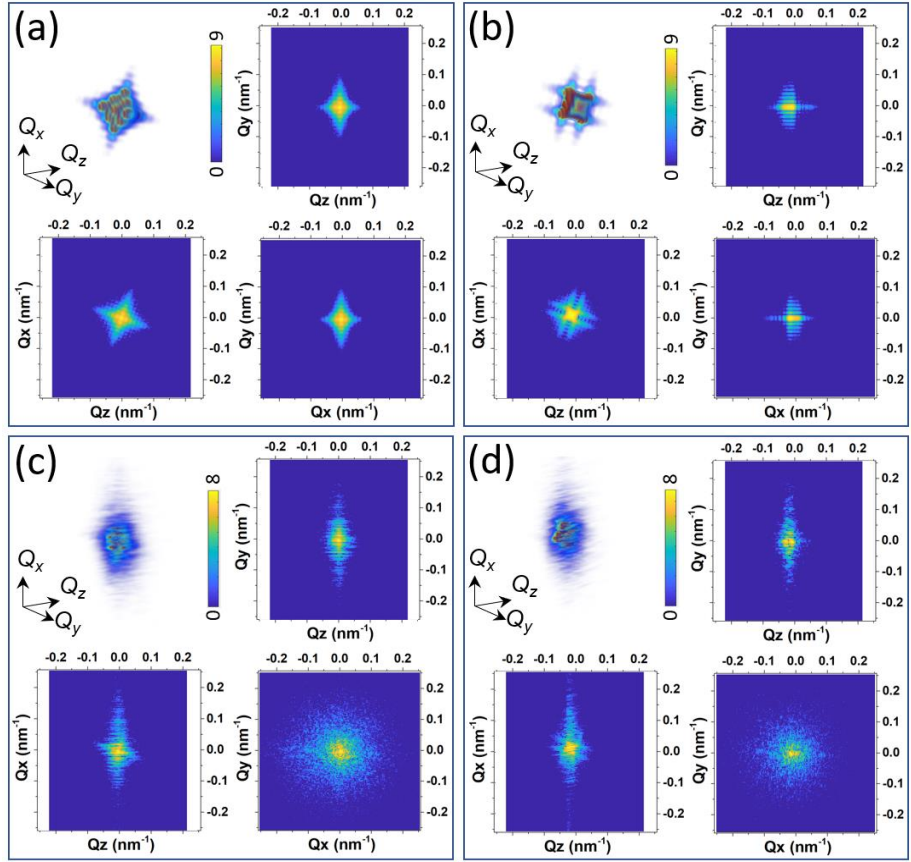

**Figure 4.** Generated BCDI and BCMI diffraction patterns. (a), (b) BCDI diffraction pattern for *model-1* and *model-2*. (c), (d) BCMI diffraction pattern for *model-1* and *model-*2. The 3D profile (top-left) and central cross-sections are shown in each panel on a logarithmic scale.

All the simulated diffraction intensities were then reconstructed 10 times using the corresponding dedicated phase retrieval algorithms (see Appendix A). In each BCMI reconstruction, a total of 800 iterations were run with different random initial object estimates and a cubic support function set to a size of $65 \times 65 \times 65$ voxels. The modulator function is assumed known in the reconstruction process. The implementation of the real-space support constraint was the same as in the HIO algorithm (Fienup, 1978, Fienup, 1982). The support function update was activated at the 200[th] iteration and performed with the shrink-wrap algorithm (Marchesini *et al.*, 2003) every 50 iterations by using a Gaussian blurring function with a sigma of 1 voxel and a 10% cutoff threshold. For each BCDI reconstruction the conventional 3D FFT-based phase retrieval algorithm was used with the same procedures of applying the real-space constraint and the support function update as in BCMI reconstruction, and ran a total number of 1500 iterations. We adopted the reciprocal space $\chi^2$ error metric (Fienup, 1997) to evaluate the reconstructions, which was defined as

$$\chi^2 = \frac{\sum_j \left\| |E_j| - \sqrt{I_j} \right\|^2}{\sum_j I_j}. \qquad (17)$$

Using the above reconstruction procedure, the BCDI reconstruction of the highly strained crystal models, unfortunately, did not produce any meaningful results. The reason is that at early iterations the reconstructed object amplitude will shrink into regions where the phase changes rapidly. This shrunken amplitude directly results in an incorrect update of the support function and eventually leads to a failed reconstruction. This phenomenon persists when using alternating HIO and error reduction (ER) algorithms. Therefore, BCDI reconstructions with a fixed tight support (one voxel larger in each dimension than the object size) were performed instead to avoid an incorrect update of the support function. The overall reconstructed objects are illustrated in Fig. 11 in Appendix B.

Figs. 5(a) and 5(b) show the BCDI reconstruction results of these two crystal models with the lowest $\chi^2$ value. The features of the object now are visible when the tight support constraint is enforced, but they still have a certain deviation from the ground truth. This best solution of crystal *model-1* is actually its conjugate, and the nonphysical 'gaps' in the reconstructed amplitude is obvious, similar to what has been reported in previous BCDI studies (Huang *et al.*, 2011, Newton, 2012, Newton *et al.*, 2010). The best reconstruction for *model-2* shown in Fig. 5(b) appears to be reconstructed to a similar, although reduced, quality when compared to the BCMI reconstruction in Fig. 5(d). When we examine all BCDI reconstructed results for this model, shown in Fig. 11 (c), it can be seen that the reconstruction shown in Fig. 5(b) is the only one which shows a resemblance to the object structure, indicating the conventional BCDI reconstruction approach for this data is not reliable. Combining the results in Fig. 11 and the $\chi^2$ curve shown in Figs. 6(a) and 6(b), we also see that the BCDI reconstructions are very sensitive to the input object estimate and convergence to a solution where the structure matches the ground truth could not be guaranteed. The BCDI reconstruction success ratio is still very limited even under a tight support constraint.

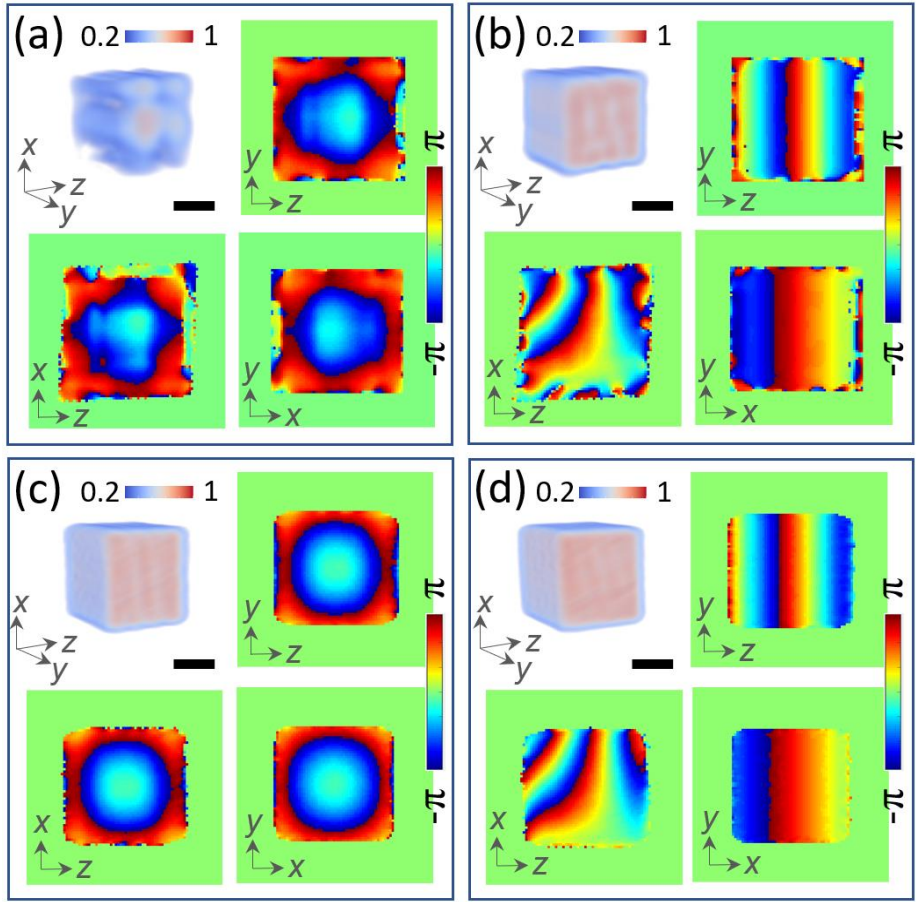

**Figure 5.** BCDI and BCMI reconstruction results for nanocrystal *model-1* and *model-2*. (a), (b) The 3D amplitude and 2D central phase cross-sections of the BCDI reconstruction results with the lowest $\chi^2$ when the tight support was used in phase retrieval. Panel (a) displays the results after performing the conjugation operation. (c), (d) The BCMI reconstruction results. All phase offsets of the results have been corrected for better comparison with the ground truth. The scale bar is 200 nm.

The BCMI reconstructions, on the contrary, show great success as the structure of the reconstructed object amplitudes and phase are highly consistent with the ground truth [see Figs. 5(c) and 5(d), and the results in Fig. 11]. All the reconstructions also show a fast convergence and the $\chi^2$ can always reach a low value below 0.07 [see Figs. 6(c) and 6(d)]. The object function, in principle, can almost be recovered correctly after tens of iterations even with a loose initial support, and the following support function updating would not encounter any difficulties as with BCDI. Moreover, no ambiguous solution appears. The reason for suppressing the ambiguous solutions in such a phase retrieval algorithm can be found in *Ref.* (Zhang *et al.*, 2016). These phenomena successfully demonstrate the robustness and accuracy of using BCMI to do the phase retrieval for highly strained nanocrystals.

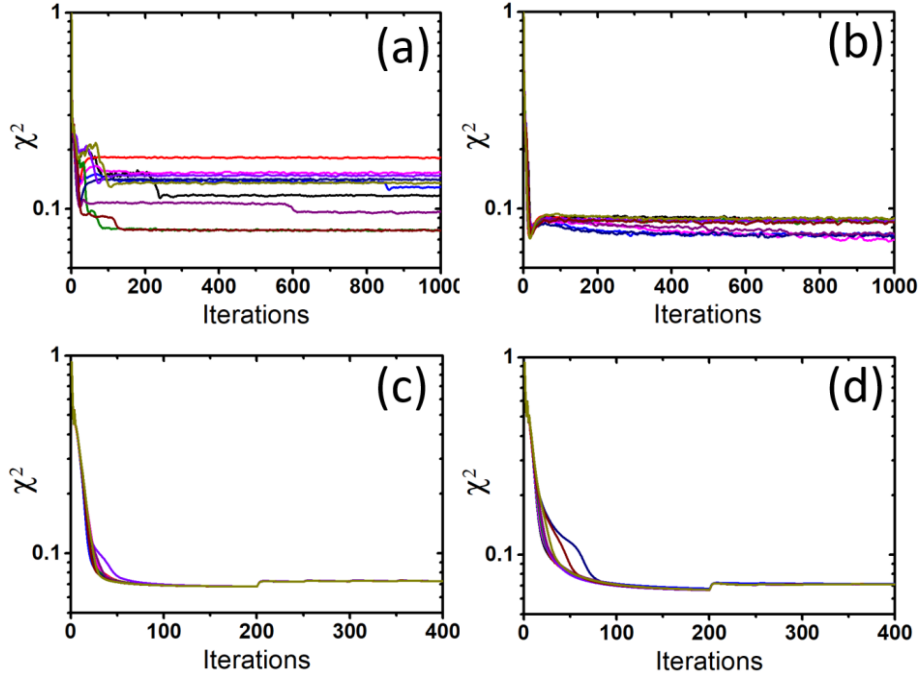

**Figure 6.** (a), (b) The $\chi^2$ curves for 10 BCDI reconstructions of the *model-1* and *model-2* crystals. (c), (d) The $\chi^2$ curves for 10 BCMI reconstructions of the *model-1* and *model-2* crystals. Only limited iterations are displayed due to no further changes of $\chi^2$. The $\chi^2$ value jump at the 200$^{th}$ iteration in (c) and (d) is caused by the activation of the shrink-wrap algorithm.

We further investigated the performance of the BCMI reconstruction for objects with different lattice strain strengths. The strain states are realized by adjusting the $\beta$ factor in the crystal phase function, where $\beta = 4, 8, 12, 16$ were used for *model-1* and $\beta = 10, 20, 30, 40$ were used for *model-2*. The generation of BCMI diffraction patterns and phase retrieval have the same procedures as before. From the reconstruction results in Figs. 7(a) and 7(c), the phase profiles for both *model-1* and *model-2* have been well reconstructed to a certain extent. However, the BCMI reconstruction is phase-structure-dependent; the reconstruction of crystal *model-1* and *model-2* struggled at different phase ranges, which is about 8 radians ($\beta = 8$) in *model-1* and 30 radians ($\beta = 30$) in *model-2*. At the same time, the BCMI reconstruction is still phase-strength-dependent, which means the higher the phase, the harder it is to converge [see Figs. 7(b) and 7(d)]. We attribute the reason for the struggled BCMI reconstruction at higher phase strength to the lack of modulation effects along the $Q_z$ direction, as a 2D wavefront modulator only provides modulation in the $Q_x Q_y$ plane. So, introducing modulation along the $Q_z$-axis, such as through the use of structured illumination, would be our next step to further improve the performance of BCMI.

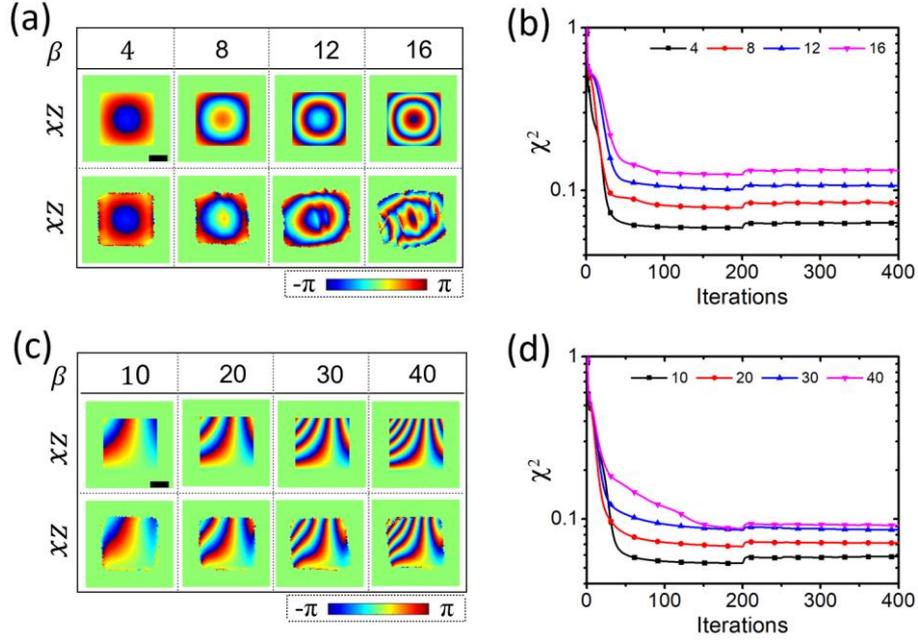

**Figure 7**. (a), (c) The BCMI reconstruction results when crystal *model-1* (a) and *model-2* (c) have different phase strengths. Only the $xz$ phase cross-section of the ground truth (first row) and the reconstructed one (second row) are shown for comparison. The phase offsets have been corrected here. The scale bar is 200 nm. (b), (d) The evolution of $\chi^2$ for the crystal *model-1* (b) and *model-2* (d) with different phase strengths. Only the first 400 iterations are shown.

## 4. Discussion of the BCMI experimental implementation

Our numerical simulations demonstrate that the proposed BCMI scheme performs well on highly strained nanocrystals. However, the experimental implementation of BCMI could pose certain challenges. The first one is to fabricate a proper wavefront modulator based on the X-ray energy used. This requires an appropriate selection of material and optimized design of the modulation function. A material that minimally absorbs X-rays while causing a larger phase shift is preferred. Additionally, it must be resistant to radiation damage and be mechanically stiff. For such purposes, metal elements with high atomic numbers and scattering factors such as tungsten, platinum, and gold are suitable candidates for hard X-ray energies. The design should follow the principle of providing effective modulation, *i.e.,* maximizing the interaction between the object scattering wave and modulator while keeping the diffracted wave within the detection area. The design of binary modulation features with random distributions is one option due to its high efficiency and ease of fabrication. Our previous CMI experiment (Zhang *et al.*, 2016) has demonstrated the success of such a design.

The second challenge could be the installation and in-place calibration of the modulator (Wang*, et al.*, 2020, Wang *et al.*, 2021). To achieve a considerable wavefront modulation, it is recommended to position the modulator in the proximity of the sample, typically at a distance of a few millimeters downstream. Such a small separation requires careful alignment and attention to avoid any collision of the modulator with the sample stage. Object-to-modulator distances as small as 1 mm in 2D transmission (Zhang *et al.*, 2016) and 2.2 mm in reflection geometry (Verezhak *et al.*, 2021) have been demonstrated. The modulator calibration using ptychography provides reliable knowledge of its transmission function as well as the object-to-modulator distance. However, it needs to be performed under the same Bragg condition. These requirements mechanically need an additional high resolution motor assembly containing three-axis translation and one-axis rotation stage. The synchronization of the motor control and ptychography data recording is also desirable.

A low-resolution calibrated modulator function could be a reason for reduced BCMI reconstruction quality. Using the same BCMI diffraction pattern as shown in Fig. 4(c), we simulated the influence of a low-resolution (blurred) modulator function on the phase retrieval process. Blurred modulator functions were generated by Gaussian smoothing the original with a constant kernel size of $5 \times 5$ pixels$^2$ but with different standard deviations of $\sigma = 1$ and $\sigma = 3$. The BCMI reconstruction procedures are the same as before and the reconstructed results are shown in Fig. 8. These results indicate that if the low-resolution modulator function was used in the phase retrieval algorithm, a distorted BCMI reconstruction would be obtained. Therefore, accurate calibration of the modulator function is the priority for BCMI experiments. Thanks to the rapid improvements in ptychography algorithms (Thibault & Menzel, 2013, Zhang *et al.*, 2013, Odstrcil *et al.*, 2018, Maiden *et al.*, 2017) and improved endstation mechanical stability (Leake *et al.*, 2019, Richard *et al.*, 2022, Carbone *et al.*, 2022), the calibration of the modulator can now be performed reliably and should not be an issue in a BCMI experiment.

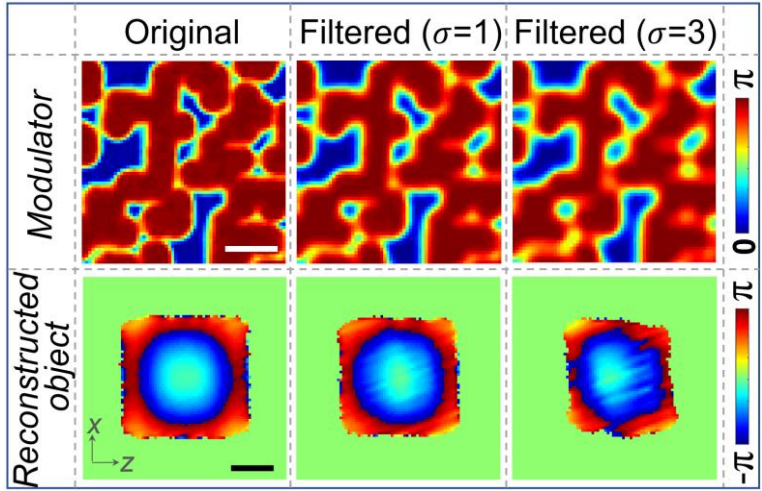

**Figure 8.** BCMI reconstruction results with a low-resolution calibrated modulator function. Only the reconstructed $xz$ phase cross-sections are shown and the phase offsets have been corrected for better comparison. The scale bar is 200 nm.

The vibration of a modulator during BCMI data acquisition could also cause a problem for the following phase retrieval. Taking the crystal *model-1* ($\beta = 6$) as an example, the impacts of the modulator vibration on BCMI were investigated. The vibration was realized by randomly shifting the modulator at each rocking angle. Here, normal-distributed $x$- and $y$-axis shifting with the standard deviation of $\sigma = 24.3$ nm, $\sigma = 36.4$ nm, and $\sigma = 48.5$ nm was considered [see Figs. 9 (a-c)]. As a result, blurred diffraction patterns that positively related to the vibration amplitude were obtained (not shown). The corresponding BCMI reconstruction performed with the same phase retrieval procedure as before are shown in Figs. 9 (d-f). It indicates that if the main vibration range does not exceed 60 nm, which is close to half the minimum size of the modulator unit, the reconstruction result is acceptable. Otherwise, the reconstructed object function could undergo degradation or distortion. This conclusion actually provides a reference for us to stabilize the modulator within some vibration range when implementing BCMI, and needs further investigation in the future. To alleviate the modulator vibration in an experiment, one can mount the modulator and the sample on the same granite table, or utilize a positive position feedback system to stabilize the modulator with high accuracy. The use of a position alignment algorithm (Zhang *et al.*, 2013) in phase retrieval could also be a method to eliminate the impact of modulator vibration on BCMI reconstruction.

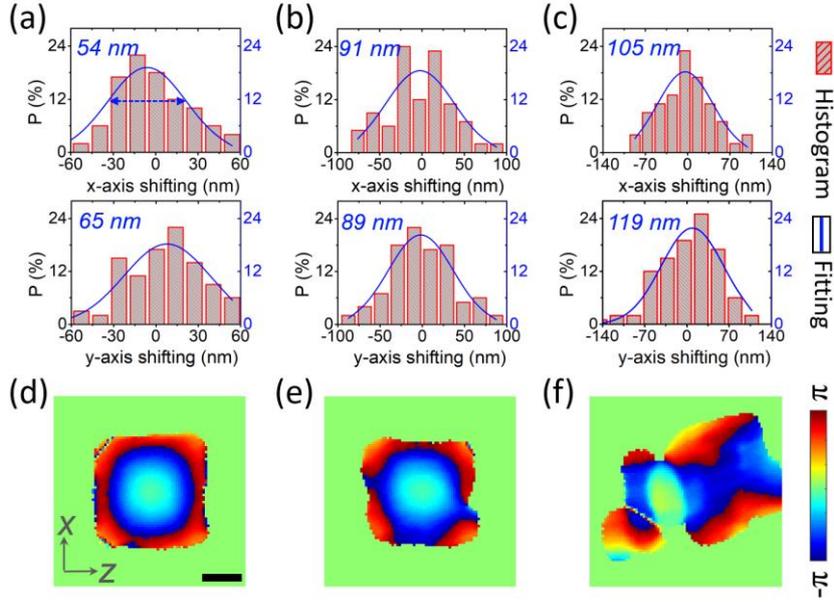

**Figure 9.** Influence of modulator vibration on BCMI reconstructions. (a-c) Simulated modulator vibration has a normal-distributed $x$- and $y$-axis shifting with the standard deviation of $\sigma = 24.3$ nm (a), $\sigma = 36.4$ nm (b), and $\sigma = 48.5$ nm (c). The FWHM of the shifting distributions are shown in the top-left corner of the plots. (d-f) The corresponding BCMI reconstruction results with respect to the modulator vibration amplitudes in (a-c). Only the reconstructed $xz$ phase cross sections are shown. The scale bar is 200 nm.

## 5 Conclusions

In summary, we have implemented wavefront modulation into the Bragg condition for imaging the 3D lattice strain of highly strained nanocrystals. The principle of BCMI has been established, which consists of a slice-by-slice model for wave propagation and a dedicated phase retrieval algorithm. This new method exhibits superior reconstruction performances than conventional BCDI in terms of fast convergence and reconstruction accuracy. In general, BCMI phase retrieval is less sensitive to the initial guesses and can suppress ambiguous solutions. The experimental implementation of the proposed BCMI method is possible thanks to the availability and reliable calibration procedure of the modulator by ptychography. The realization of this method will provide a powerful tool for measuring the lattice strain of the monodispersed nanocrystals, and particularly beneficial to research related to high-pressure and ferroelectric materials.

**APPENDIX A: BCMI Phase Retrieval Algorithm**

Following the BCMI propagation models, a dedicated phase retrieval algorithm was composed and its pseudocode is given in Fig. 10. The inputs of this algorithm are diffraction intensity measurements $\{I_j\}_{j=1}^{J}$, the prior knowledge of the modulator function $Modu$, the support function $supp$ and the object estimate $S^0(r)$. To speed up the iterations, the $Q_z^j$-related phase term $\{\phi_j\}_{j=1}^{J}$ [in equation (3)] can also be calculated in advance for use. In the main iteration cycle, all the recorded diffraction intensity slices, $I_j$, will be looped through to update the running estimate of real-space object representation $S(r)$. For each slice, the corresponding object projection is propagated to the detector plane using the forward propagation model of equations (6)-(10) (steps 2-5 in Fig. 10). The diffraction wave estimate $E_j$ is then updated by applying the modulus constraint defined as

$$\hat{E}_j = \sqrt{I_j} \frac{E_j}{|E_j|}. \tag{17}$$

Using $\hat{E}_j$, the real-space object component $S_j(r)$ can be obtained using the BCMI backward propagation model described by equations (11)-(14), and the updated object $\hat{S}(r)$ is then calculated by summing all the $S_j(r)$ with equation (5) (steps 6-8 in Fig. 10). After getting $\hat{S}(r)$ in one loop, the real-space support constraint is applied (step 9 in Fig. 10). If necessary, the support function can be refined using the shrink-wrap algorithm after a certain number of iterations (step 10 in Fig. 10).

```
// Given quantities
{I_j}_{j=1}^{J} (intensity measurements); Modu (modulator function); supp (support
function); S^0 (initial object estimate); {φ_j}_{j=1}^{J} (the Q_z^j-related phase term)

for k = 1 ... do  // The main iteration
    // ① Give the current estimate
    S^k = S^{k-1}
    Ŝ^k = 0

    for j = 1...J   //loop each 2D diffraction pattern
        // ② Generate the object projection along z-axis
        S_j^k = ∫ S^k · φ_j dz
        // ③ Propagate the object projection to modulator plane
        E_j^k(m) = P{S_j^k}
        // ④ Apply modulation
        E_j^k(M) = Modu · E_j^k(m)
        // ⑤ Propagate to detector plane
        E_j^k(D) = F{E_j^k(M)}
        // ⑥ Apply the modulus constraint and propagate back to the modulator plane
        Ê_j^k(M) = F^{-1}{ √I_j · E_j^k(D)/|E_j^k(D)| }
        // ⑦ Undo modulation
        Ê_j^k(m) = E_j^k(m) + Modu*/max(|Modu|^2) · (Ê_j^k(M) − E_j^k(M))
        // ⑧ Propagate back to the object plane to update the object
        Ŝ^k += φ_j* · P^{-1}{Ê_j^k(m)}
    end

    // ⑨ Do support constraint
    if ER type
        S^k = Ŝ^k · supp
    elseif HIO type
        S^k = Ŝ^k · supp + β(S^k − Ŝ^k) · (1 − supp)
    end

    // ⑩ Update the support in each N^{th} iteration
    if k is a multiple of N; supp = Shrink_wrap(S^k); end
end
```

**Figure 10.** Pseudocode of the BCMI phase retrieval algorithm. The pair $\mathcal{P}\{\cdot\}$ and $\mathcal{P}^{-1}\{\cdot\}$ are the forward and backward propagation operators between the object and modulator plane. The pair $\mathcal{F}\{\cdot\}$ and $\mathcal{F}^{-1}\{\cdot\}$ are the direct and inverse 2D Fourier transforms. The waves in the front and rear plane of the modulator, and in the detector plane, are marked by m, M, and D, respectively.

## APPENDIX B: Overall BCDI and BCMI Reconstruction Results of *model-1* and *mode-2* Crystals

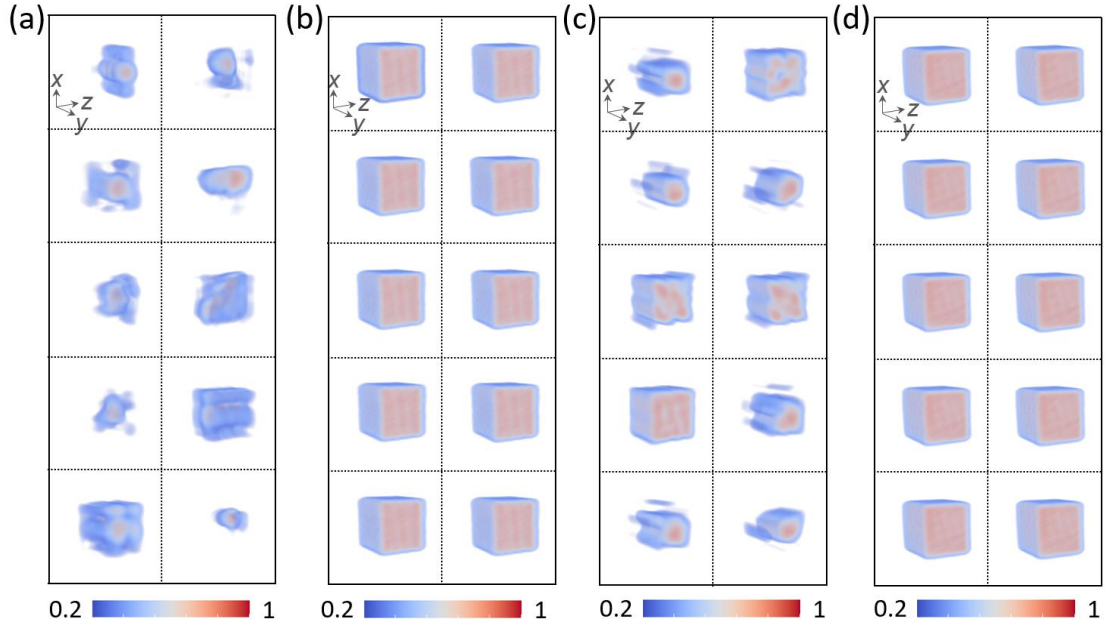

**Figure 11.** (a), (b) 10 BCDI and 10 BCMI reconstruction results of crystal *model-1*. (c), (d) 10 BCDI and 10 BCMI reconstruction results of crystal *model-2*. Only the 3D amplitudes are shown.

**Acknowledgment**

We thank Dr. Peng Li for his helpful suggestions and Dr. Gerard Hinsley for his constructive criticism and careful reading of the manuscript. This work was funded in part by the National Natural Science Foundation of China (12074167, 11775105), Shenzhen Science and Technology Program (KQTD20170810110313773), Centers for Mechanical Engineering Research and Education at MIT and SUSTech (6941806), and Postdoctoral Research Foundation of China (2021M701566).

**Disclosures**

The authors declare no conflicts of interest.

**References**

Carbone, D., Kalbfleisch, S., Johansson, U., Björling, A., Kahnt, M., Sala, S., Stankevic, T., Rodriguez-Fernandez, A., Bring, B., Matej, Z., Bell, P., Erb, D., Hardion, V., Weninger, C., Al-Sallami, H., Lidon-Simon, J., Carlson, S., Jerrebo, A., Norsk Jensen, B., Bjermo, A., Åhnberg, K. & Roslund, L. (2022). *Journal of Synchrotron Radiation* **29**, 876-887.

Carnis, J., Kshirsagar, A. R., Wu, L., Dupraz, M., Labat, S., Texier, M., Favre, L., Gao, L., Oropeza, F. E., Gazit, N., Almog, E., Campos, A., Micha, J. S., Hensen, E. J. M., Leake, S. J., Schulli, T. U.,

Rabkin, E., Thomas, O., Poloni, R., Hofmann, J. P. & Richard, M. I. (2021). *Nat Commun* **12**, 5385.

Cha, W., Ulvestad, A., Allain, M., Chamard, V., Harder, R., Leake, S. J., Maser, J., Fuoss, P. H. & Hruszkewycz, S. O. (2016). *Phys Rev Lett* **117**, 225501.

Clark, J. N., Ihli, J., Schenk, A. S., Kim, Y. Y., Kulak, A. N., Campbell, J. M., Nisbet, G., Meldrum, F. C. & Robinson, I. K. (2015). *Nat Mater* **14**, 780-784.

Dupraz, M., Li, N., Carnis, J., Wu, L., Labat, S., Chatelier, C., van de Poll, R., Hofmann, J. P., Almog, E., Leake, S. J., Watier, Y., Lazarev, S., Westermeier, F., Sprung, M., Hensen, E. J. M., Thomas, O., Rabkin, E. & Richard, M. I. (2022). *Nat Commun* **13**, 3003.

Favre-Nicolin, V., Mastropietro, F., Eymery, J., Camacho, D., Niquet, Y. M., Borg, B. M., Messing, M. E., Wernersson, L. E., Algra, R. E., Bakkers, E. P. A. M., Metzger, T. H., Harder, R. & Robinson, I. K. (2010). *New Journal of Physics* **12**.

Fienup, J. R. (1978). *OPTICS LETTERS* **3**, 27-29.

Fienup, J. R. (1982). *APPLIED OPTICS* **Vol. 21**, 2758-2769.

Fienup, J. R. (1997). *Applied Optics* **36**, 8352-8357.

Gao, Y., Huang, X., Yan, H. & Williams, G. J. (2021). *Physical Review B* **103**, 014102.

Hruszkewycz, S. O., Allain, M., Holt, M. V., Murray, C. E., Holt, J. R., Fuoss, P. H. & Chamard, V. (2017). *Nat Mater* **16**, 244-251.

Huang, X., Harder, R., Xiong, G., Shi, X. & Robinson, I. (2011). *Physical Review B* **83**.

Karpov, D., Liu, Z., Rolo, T. D. S., Harder, R., Balachandran, P. V., Xue, D., Lookman, T. & Fohtung, E. (2017). *Nat Commun* **8**, 280.

Kawaguchi, T., Keller, T. F., Runge, H., Gelisio, L., Seitz, C., Kim, Y. Y., Maxey, E. R., Cha, W., Ulvestad, A., Hruszkewycz, S. O., Harder, R., Vartanyants, I. A., Stierle, A. & You, H. (2019). *Phys Rev Lett* **123**, 246001.

Lauraux, F., Labat, S., Yehya, S., Richard, M.-I., Leake, S. J., Zhou, T., Micha, J.-S., Robach, O., Kovalenko, O., Rabkin, E., Schülli, T. U., Thomas, O. & Cornelius, T. W. (2021). *Crystals* **11**.

Leake, S. J., Chahine, G. A., Djazouli, H., Zhou, T., Richter, C., Hilhorst, J., Petit, L., Richard, M. I., Morawe, C., Barrett, R., Zhang, L., Homs-Regojo, R. A., Favre-Nicolin, V., Boesecke, P. & Schulli, T. U. (2019). *J Synchrotron Radiat* **26**, 571-584.

Li, P., Maddali, S., Pateras, A., Calvo-Almazan, I., Hruszkewycz, S. O., Cha, W., Chamard, V. & Allain, M. (2020). *Journal of Applied Crystallography* **53**, 404-418.

Li, P., Phillips, N. W., Leake, S., Allain, M., Hofmann, F. & Chamard, V. (2021). *Nat Commun* **12**, 7059.

Maiden, A., Johnson, D. & Li, P. (2017). *Optica* **4**, 736.

Marchesini, S., He, H., Chapman, H. N., Hau-Riege, S. P., Noy, A., Howells, M. R., Weierstall, U. & Spence, J. C. H. (2003). *Physical Review B* **68**.

Minkevich, A. A., Baumbach, T., Gailhanou, M. & Thomas, O. (2008). *Physical Review B* **78**.

Minkevich, A. A., Fohtung, E., Slobodskyy, T., Riotte, M., Grigoriev, D., Schmidbauer, M., Irvine, A. C., Novák, V., Holý, V. & Baumbach, T. (2011). *Physical Review B* **84**.

Newton, M. C. (2012). *Phys Rev E Stat Nonlin Soft Matter Phys* **85**, 056706.

Newton, M. C., Harder, R., Huang, X., Xiong, G. & Robinson, I. K. (2010). *Physical Review B* **82**.

Odstrcil, M., Menzel, A. & Guizar-Sicairos, M. (2018). *Opt Express* **26**, 3108-3123.

Pfeifer, M. A., Williams, G. J., Vartanyants, I. A., Harder, R. & Robinson, I. K. (2006). *Nature* **442**, 63-66.

Richard, M. I., Labat, S., Dupraz, M., Li, N., Bellec, E., Boesecke, P., Djazouli, H., Eymery, J., Thomas,

O., Schulli, T. U., Santala, M. K. & Leake, S. J. (2022). *J Appl Crystallogr* **55**, 621-625.

Robinson, I. & Harder, R. (2009). *Nat Mater* **8**, 291-298.

Robinson, I. K., Vartanyants, I. A., Williams, G. J., Pfeifer, M. A. & Pitney, J. A. (2001). *Phys Rev Lett* **87**, 195505.

Rodenburg, J. M. & Faulkner, H. M. L. (2004). *Applied Physics Letters* **85**, 4795-4797.

Singer, A., Zhang, M., Hy, S., Cela, D., Fang, C., Wynn, T. A., Qiu, B., Xia, Y., Liu, Z., Ulvestad, A., Hua, N., Wingert, J., Liu, H., Sprung, M., Zozulya, A. V., Maxey, E., Harder, R., Meng, Y. S. & Shpyrko, O. G. (2018). *Nature Energy* **3**, 641-647.

Thibault, P. & Menzel, A. (2013). *Nature* **494**, 68-71.

Ulvestad, A., Cha, W., Calvo-Almazan, I., Maddali, S., Wild, S. M., Maxey, E., Dupraz, M. & Hruszkewycz, S. O. (2018). *arXiv preprint arXiv* **1808**.

Vartanyants, I. A. & Robinson, I. K. (2001). *J. Phys.: Condens. Matter* 10593–10611.

Verezhak, M., Van Petegem, S., Rodriguez-Fernandez, A., Godard, P., Wakonig, K., Karpov, D., Jacques, V. L. R., Menzel, A., Thilly, L. & Diaz, A. (2021). *Physical Review B* **103**.

Wang, B., He, Z. & Zhang, F. (2021). *Opt Express* **29**, 30035-30044.

Wang, B., Wang, Q., Lyu, W. & Zhang, F. (2020). *Ultramicroscopy* **216**, 113034.

Wang, Z., Gorobtsov, O. & Singer, A. (2020). *New Journal of Physics* **22**, 013021.

Yang, W., Huang, X., Harder, R., Clark, J. N., Robinson, I. K. & Mao, H. K. (2013). *Nat Commun* **4**, 1680.

Zhang, F., Chen, B., Morrison, G. R., Vila-Comamala, J., Guizar-Sicairos, M. & Robinson, I. K. (2016). *Nat Commun* **7**, 13367.

Zhang, F., Peterson, I., Vila-Comamala, J., Diaz, A., Berenguer, F., Bean, R., Chen, B., Menzel, A., Robinson, I. K. & Rodenburg, J. M. (2013). *Opt Express* **21**, 13592-13606.

Zhang, F. & Rodenburg, J. M. (2010). *Physical Review B* **82**.